\documentclass[10pt, conference, compsocconf]{IEEEtran}
\usepackage{amsmath}
\usepackage{amsthm}
\usepackage{graphicx}
\usepackage{subfigure}
\usepackage{mathrsfs}
\usepackage{algorithm}
\usepackage{algorithmic}
\usepackage{color}
\usepackage{enumerate}
\usepackage{booktabs}
\usepackage{amsmath}
\usepackage{amssymb}
\usepackage{multirow}
\usepackage[english, greek]{babel}

\begin{document} 
%
\IEEEoverridecommandlockouts
\selectlanguage{english}

\title{\huge LazyCtrl: Scalable Network Control for Cloud Data Centers}



%

\author{
\IEEEauthorblockN{
Kai Zheng\IEEEauthorrefmark{1},
Lin Wang\IEEEauthorrefmark{2},
Baohua Yang\IEEEauthorrefmark{1},
Yi Sun\IEEEauthorrefmark{2},
Yue Zhang\IEEEauthorrefmark{1},
Steve Uhlig\IEEEauthorrefmark{3}
}
\IEEEauthorblockA{
\IEEEauthorrefmark{1}IBM Research} 
\IEEEauthorblockA{
\IEEEauthorrefmark{2}Institute of Computing Technology, Chinese Academy of Sciences} 
\IEEEauthorblockA{
\IEEEauthorrefmark{3}Queen Mary University of London} 
}


\maketitle

\begin{abstract}
The advent of software defined networking enables flexible, reliable and feature-rich control planes for data center networks. However, the tight coupling of centralized control and complete visibility leads to a wide range of issues among which scalability has risen to prominence. To address this, we present LazyCtrl, a novel hybrid control plane design for data center networks where network control is carried out by distributed control mechanisms inside independent groups of switches while complemented with a global controller. Our design is motivated by the observation that data center traffic is usually highly skewed and thus edge switches can be grouped according to traffic locality. LazyCtrl aims at bringing laziness to the global controller by dynamically devolving most of the control tasks to independent switch groups to process frequent intra-group events near datapaths while handling rare inter-group or other specified events by the controller. We implement LazyCtrl and build a prototype based on Open vSwich and Floodlight. Trace-driven experiments on our prototype show that an effective switch grouping is easy to maintain in multi-tenant clouds and the central controller can be significantly shielded by staying lazy, with its workload reduced by up to 82\%.

\end{abstract}

\section{Introduction}

Public clouds are becoming increasingly popular due to their \emph{pay-as-you-go} model, which attracts many small and medium business. Some of them, thanks to their success, have grown very large, each containing hundreds thousand of servers and hosting up to millions of virtual machines \cite{EC2}. To support flexible and efficient inter-node communication in these large-scale cloud data centers, researchers have proposed many novel designs (\emph{e.g.}, \cite{Al-Fares_Loukissas-2008, Guo_Lu_Li-2009}) for data center networks to replace traditional tree-based architectures. However, the routing and forwarding protocols used in most designs are restricted to very specific deployment settings, leading to inflexible configuration and management. The situation has been revolutionized by Software Defined Networking (SDN), where the control plane, separated from the data plane, is implemented with a logically centralized controller. As a result, when adopting SDN, flow-based polices can be conveniently applied to achieve fine-grained control over the data center network.

While flow-based centralized control has been recently employed in several proposals for traffic management in data center networks \cite{Al-Fares_Radhakrishnan-2010, Heller_Seetharaman-2010, Benson_Anand-2011}, the excessive coupling of central control and complete visibility has brought many scalability challenges to both the network control and data planes in large-scale data centers. On the one hand, having the controller to set up all flows would bring too much workload to the controller and such centralized bottlenecks are difficult to scale. On the other hand, maintaining visibility of all flows in a large-scale network can require hundreds of thousands of flow table entries at each switch, which is far from practical for commodity switches.

\subsection{Bringing Laziness  to the Controller}

\begin{center}
``\emph{Laziness is the first step towards efficiency.}''

\hspace{3.6cm}{-- Patrick Bennett}
\end{center}
It has been demonstrated that full control and visibility over all flows are not always necessary and devolving some control authority to the data plane by proactively suppressing frequent events can result in better scalability in software defined data center networks \cite{Curtis_Mogul-2011}. However, the right granularity of flows to be handled by the controller is still not clear (or hard to define). In this paper, we advocate a new solution for control devolvement in data center networks based on traffic locality. Our idea stems from the observation that traffic distribution in data centers (especially those with multi-tenancy support) could be highly skewed, \emph{i.e.}, frequent communications are more likely to take place inside certain small groups of hosts. As a result, it is possible to shield the global controller from many frequent events inside these groups if distributed control mechanism is applied independently in each of the groups.

We propose LazyCtrl, a hybrid network control plane design for large-scale data centers, which seeks to bring \emph{laziness} to the global controller. In the LazyCtrl design, edge switches are grouped dynamically according to their communication affinity. The central controller devolves the coarse-grained control for frequent intra-group events to each switch group while handling infrequent inter-group and other specified (fine-grained) control tasks by itself. Each switch group autonomously carries out \emph{distributed control} within the group, keeping the intra-group packets in the data plane. The controller groups the switches in such a way that the size of each group is as large as possible to exhaust switches' memory (such as TCAMs) capacity while inter-group traffic is minimized to support the laziness of the controller.

We have completed a full implementation of LazyCtrl based on Open vSwitch and the Floodlight OpenFlow controller. Experiments on our prototype with both real and synthetic traffic traces show that an effective switch grouping is easy to maintain in multi-tenant clouds and the hybrid control design highly reduces the workload of the controller and provides lower delay in packet forwarding. As expected, the laziness we introduced to the controller decouples centralized control and complete visibility and consequently scale the system much better compared with totally centralized designs.

Section~\ref{sec:motivation} reveals some observations that motivate our design. Section~\ref{sec:design} presents the LazyCtrl architecture with design details. Section~\ref{sec:implementation} presents our implementation, followed by the performance evaluation in Section~\ref{sec:evaluation}. Section~\ref{sec:conclusion} concludes the paper.

\subsection{Related Work}

Ethernet stands as one of the most widely used networking technologies today due to its \emph{plug-and-play} semantics such as automatic host location learning and flat addressing, which can highly simplify many aspects of network configuration and ensure service continuity. However, replying on network-wide dissemination of per-host information makes Ethernet-based solutions difficult to scale and forcing paths to comprise a spanning tree introduces substantial inefficiencies. In contrast, IP networks can easily scale to large networks but require massive effort to configure and manage.

As a promising solution for building large-scale data center networks, network overlay can exploit the advantages of both Ethernet and IP networks. An overlay network in a data center consists in creating a dynamic mapping between the overlay (virtual) network and the underlying (physical) infrastructure. This mapping ensures that packets can be transmitted by the routing substrate between any pair of overlay nodes. However, in order to handle location resolution at network edge, a global location information base has to be maintained, which can be challenging in large networks.

There has been a large body of work falling in this category. SEATTLE \cite{Kim_Caesar-2008} simplifies network management by flat addressing while providing hash-based resolution of host information (using a one-hop DHT) to ensure scalability. VL2 \cite{Greenberg_Hamilton-2009} implements a layer 2.5 stack on hosts and uses IP-in-IP encapsulation to deliver packets. PortLand \cite{Niranjan_Pamboris-2009} assigns Pseudo MAC (PMAC) addresses to all end hosts to enable efficient, provably loop-free forwarding with small switch state while leveraging a central fabric manager to address IP to PMAC translation in multi-rooted tree networks. NetLord \cite{Mudigonda_Yalagandula-2011} employs a light-weight agent in the end-host hypervisors to encapsulate and transmit packets over an underlying, multi-path L2 network, using an unusual combination of IP and Ethernet packet headers.

With the rapid evolvement of SDN, flow-based centralized control has been recently adopted as a mainstream control plane design for data center networks. As one of the first SDN solutions for enterprise networks, Ethane \cite{Casado_Freedman-2007} enables the direct application of fine-grained flow-based polices to the network by coupling flow switches with a centralized controller. However, exposing all flows to the controller could bring too much workload to the controller, leading to poor scalability. Even after applying multi-threading optimizations that help achieve graceful linear core scaling factors \cite{Tootoonchian_Gorbunov-2012}, the gap between actual and desired performance of the centralized controller is still very significant. It was shown that the popular OpenFlow controller can only be able to handle approximately 30 thousand flow initiation requests per second on commodity x86 platforms \cite{Tavakoli_Casado-2009}. Unfortunately, a small network consists of only 100 switches could have a spike of more than 10 million flow arrivals per second \cite{Benson_Akella-2010}. Even after applying multi-threading optimizations that help achieve graceful linear core scaling factors \cite{Tootoonchian_Gorbunov-2012}, the gap between actual and desired performance of the centralized controller is still very significant.

Recently, massive effort has been devoted to scaling centralized control to large networks. Onix \cite{Koponen_Casado-2010}, HyperFlow \cite{Ganjali_Tootoonchian-2010}, ElastiCon \cite{Dixit_Hao-2013}, and Pratyaastha \cite{Krishnamurthy_Chandrabose-2014} are distributed platforms on top of which the network control plane can be implemented as a distributed system. DIFANE \cite{Yu_Rexford-2010} aims at handling all traffic in the data plane by selectively directing packets through intermediate (authority) switches that store the necessary rules pre-installed by the controller. DevoFlow \cite{Curtis_Mogul-2011} decouples control and global visibility and partly devolves control to switches by employing rule cloning and local actions at switches. Kandoo \cite{Yeganeh_Ganjali-2012} is a two-layer control framework where network applications are classified and local and global control applications are handled by bottom- and top-layer controllers, respectively. Recently, Jain \emph{et al.} \cite{Jain-B4} presented B4, a private WAN connecting Google's data centers worldwide based on a multi-layer software defined networking architecture.

LazyCtrl also targets the scalability issue of centralized control in large-scale data center networks, but is orthogonal to the above designs in the sense that it employs a hybrid control model, aiming at trying best to offload frequent coarse-grained control tasks from the central controller and handle them using distributed control mechanisms near datapaths. Therefore, the aforementioned research effort for scaling flow-based fine-grained control is still applicable on top of LazyCtrl to further mitigate the performance bottleneck at the controller and consequently improve control plane scalability in data center networks.

\section{Motivation}
\label{sec:motivation}

The following salient features of current cloud data centers largely motivate our design of LazyCtrl.

\subsection{Traffic Locality in Data Centers}

In cloud data centers, the traffic among the hosts is usually unevenly distributed and is strongly localized within some groups of hosts. To verify the correctness of this notion, we collected a day-long traffic trace from a production data center in Europe running multi-tenant applications and made the following quantitative findings:
\begin{itemize}
\item[$\triangleright$] {\it The traffic distribution is uneven among hosts.} Among a total of 6509 hosts, only 11,602 of more than 20 million distinct $\langle src, dst \rangle$ host pairs exchanged traffic in the trace. And over 90\% of the flows are contributed by about 10\% of the host pairs that exchanged traffic.
\item[$\triangleright$] {\it The traffic appears to be concentrated within some groups of hosts.} For example, when partitioning the 6509 hosts evenly into 5 groups using $k$-way partitioning, we observe that only less than 9.8\% of the traffic traversed different groups. We define the \emph{centrality} of a group as the ratio (in $[0, 1]$) of the intra-group traffic and the total traffic related to the hosts in this group. For the collected trace, the average centrality of the 5 groups is 0.853, indicating a very high concentration of the data center traffic.
\end{itemize}

The above findings are not accidental and similar evidences can be found in \cite{Benson_Akella-2010, Brodersen_Scellato-2012}. Actually, in a multi-tenant data center, network traffic tends to be localized within each tenant, as the applications from different tenants are isolated by virtualization techniques \cite{Lam_Radhakrishnan-2012}. Therefore, we believe that by taking advantage of traffic locality, a global, fine-grained, and real-time network control may not be necessary for multi-tenant data centers.

\subsection{Relatively Stable Tenant Size}

For multi-tenant cloud data centers, we observe that the number of virtual machines for a single tenant is changing slightly, while the number of tenant users, as well as the total number of hosts in a multi-tenant data center, is experiencing a significant increase. For Amazon, a popular cloud service provider, the number of tenants, as well as total virtual machine instances of Amazon's EC2, grew about 2.5 times annually since 2006 \cite{Amazon-estimate}. The total number of objects held by Amazon S3 has grown 150 times from 2006 to 2011 \cite{Amazon-S3}. In contrast, the size of a specific tenant in terms of number of rented virtual machines is constantly around 20--100 \cite{EC2}. These facts consequently lead to the property that traffic is aggregated within some size-limited groups of hosts in multi-tenant data centers as the traffic exchanged among different tenant slices is very limited. By taking full advantage of this property, we show that the explosive increase in the number of tenants does not necessarily result in scalability issues for centralized control in data center networks. 

\section{Design}
\label{sec:design}

LazyCtrl realizes a hybrid control plane for data center networks. In this section, we discuss four aspects of its design: the architecture, the switch grouping scheme, the packet forwarding routine, and the failover mechanisms. We first provide a high-level overview to state the intuition of our design.

\subsection{High-level Overview}

In conventional flow-based centralized control environments such as those based on OpenFlow \cite{McKeown_Anderson-2008}, the controller maintains the network-wide state (the host-to-switch mapping here) and handles all the flows between every pair of switches that exchange data, bringing extremely high burden to the controller. LazyCtrl mitigates this problem by clustering the switches into multiple switch groups according to their communication affinity and devolving intra-group control to these switch groups (termed Local Control Group, LCG).\footnote{We will use group and local control group (LCG) interchangeably in the rest of this paper.} To support its laziness, the controller prefers clustering the switches into a few big groups in order to reduce inter-group communication. However, larger group size would result in larger distributed forwarding tables and more control tasks inside each local control group. Due to the limited size of high-speed memory in switches, the largest size of a group will be constrained by some constant. {\it The controller clusters the switches in such a way that the size of each group is maximized under a given limit while the inter-group traffic volume is minimized.}

\begin{figure}[!t]
\centering
\includegraphics[scale=0.37]{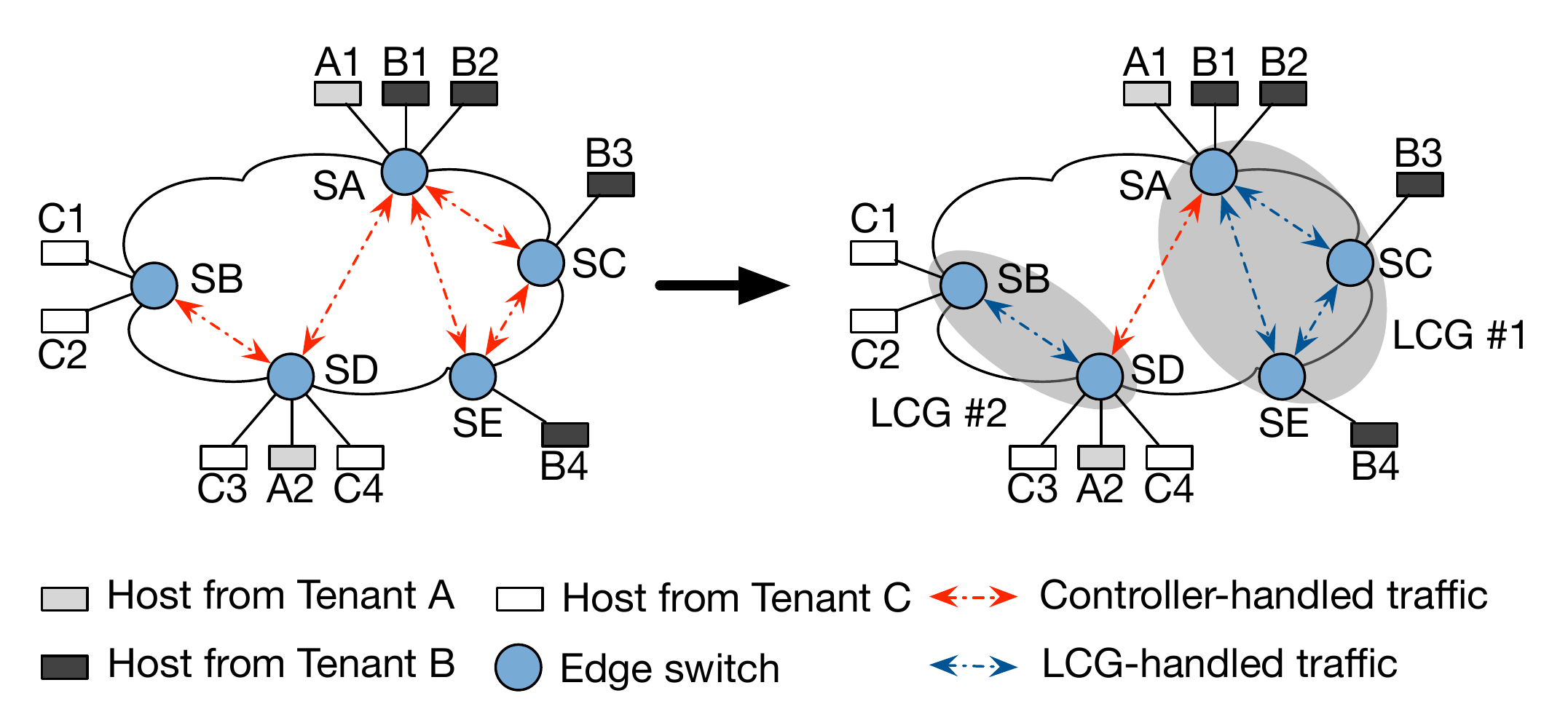}
\caption{\label{fig:example}Example to demonstrate the idea of LazyCtrl. Edge switches are clustered into multiple local control groups according to their communication affinity.}
\vspace{-0.5cm}

\end{figure}

{\bf Example:} Consider a multi-tenant cloud data center containing a central controller and five edge switches (namely \textsf{SA}, \textsf{SB}, \textsf{SC}, \textsf{SD}, and \textsf{SE}) with hosts\footnote{With a bit abuse of notation, we will use host to refer to virtual machine that is running in a physical server.} directly attached. We focus on the scenario shown in Fig.~\ref{fig:example}. There are three tenants, \textsf{A}, \textsf{B}, and \textsf{C}, each of which has some virtual machines. The left figure illustrates the case when centralized controlling is applied directly and thus the central controller has to handle all the flows among all edge switches. LazyCtrl changes this situation by clustering edge switches into independently groups. As can be seen in the right figure, the controller clusters \textsf{SA}, \textsf{SC}, and \textsf{SE} into the first group while \textsf{SB} and \textsf{SD} together form the second group. (We assume that the group size limit is three in our example.) This way, the traffic within the first group (\emph{e.g.}, \textsf{SA}$\leftrightarrow$\textsf{SC}), as well as the traffic within the second group (\emph{e.g.}, \textsf{SB}$\leftrightarrow$\textsf{SD}), can be handled by carrying out local control mechanism that is dedicated for each group. The controller then is only needed to take charge of the inter-group traffic, \emph{i.e.}, \textsf{SA}$\leftrightarrow$\textsf{SD}. The switches will be dynamically regrouped in response to traffic variation.

\subsection{LazyCtrl Architecture}

\begin{figure}[!t]
\centering
\includegraphics[scale=0.37]{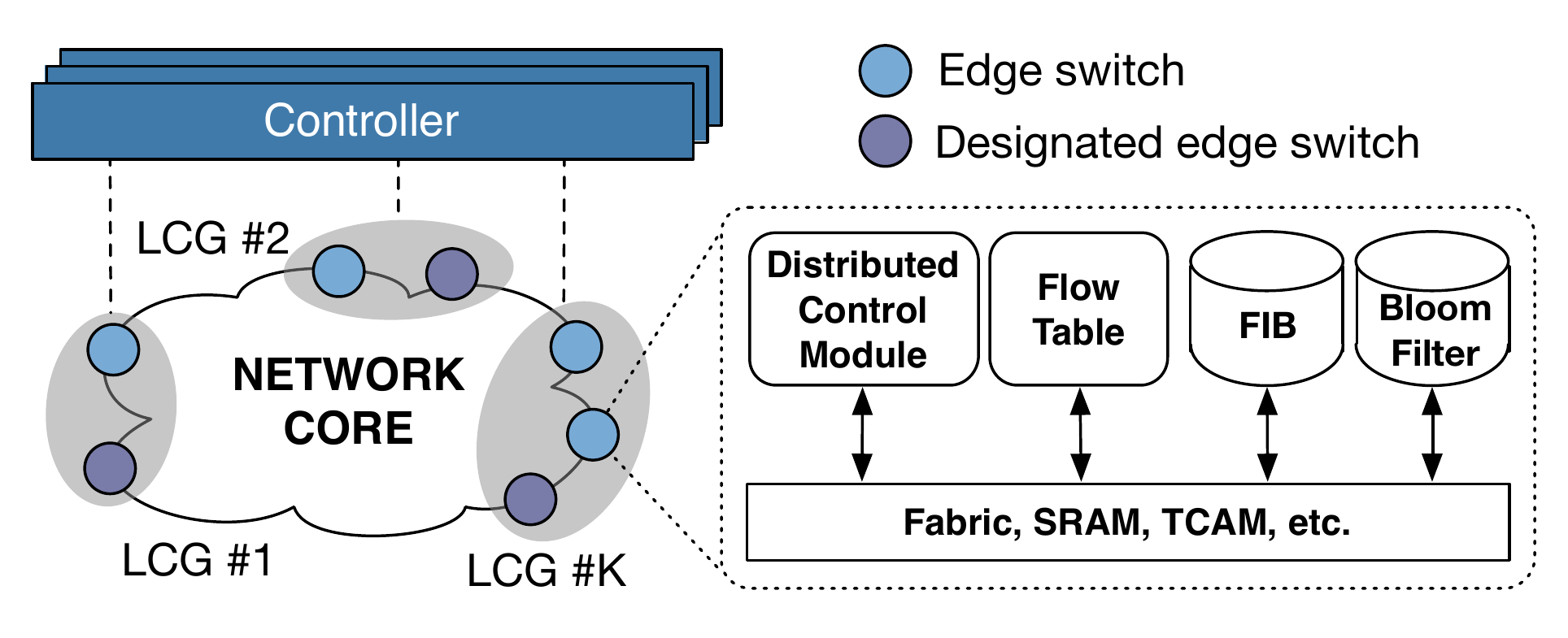}

\caption{\label{fig:arch} Architecture design of LazyCtrl, where the network control plane consists of a logically centralized controller and distributed control modules in multiple local control groups.}
\vspace{-0.5cm}

\end{figure}

The architecture design of LazyCtrl is depicted in Fig.~\ref{fig:arch}. In our design, the network is separated into two parts: the core and the edge. We employ a hybrid control model where control tasks are handled by the distributed control mechanisms in LCGs at the network edge, complemented by a central controller.

\subsubsection{Core--Edge Separation}

Our design splits the core from the edge. The network core can be any simple and scalable network (\emph{e.g.}, an IP unicast network), which serves as the underlay providing connectivity for the edge switches. The core--edge separation releases the network core from handling complicated and dynamic network control tasks (\emph{e.g.}, network virtualization, virtual machine migration) and thus allows the network core to be constrained only by performance and reliability. Since our focus is the control plane, we omit the detailed design of the network core.

In contrast, the network edge is in charge of network intelligence, \emph{i.e.}, host-to-switch mapping. The layer two virtual networks (overlays) for providing connectivity for the edge switches are conducted by the network edge via encapsulation or tunneling on top of the underlying physical network core. As a result, one-hop distance can be assumed for each pair of edge switches. We introduce a hybrid control model for the control plane to handle network control tasks. 

\subsubsection{Hybrid Control Model}

To extend the scalability of the control plane, we introduce a hybrid control model in the LazyCtrl design. This hybrid control model involves a central controller and a set of local control groups.

The central controller has holistic visibility over the entire data center network and is responsible for $\mathit{i})$ maintaining a Central Location Information Base (C-LIB) which preserves host location information, $\mathit{ii})$ adapting the grouping of the edge switches, and $\mathit{iii})$ managing the flow tables on the edge switches to handle inter-group traffic and any specific traffic that needs flexible centralized control. The goal of the central controller is to stay lazy by devolving as many control tasks as possible to the local control groups. The central controller can be a stand-alone physical server or a logical controller comprised of a cluster of servers with strong reliability and coherency of network state.

A local control group is a group of edges switches whose clients are observed to have frequent mutual communication. These switches are grouped together by the controller and share the network state with each other consistently. Each local control group employs a distributed control mechanism to take over the control workload of intra-group traffic from the controller. The distributed control mechanism inside each group is carried out by equipping each edge switch with some local forwarding tables, which are maintained by the switches themselves. These local forwarding tables keep track of network states such as host-to-switch mapping inside the corresponding group. For each local control group, a designated switch (with some backups) is selected randomly by the controller, which is responsible for aggregating group-wide network states from the edge switches in this group and reporting them to the controller in an asynchronous manner. 

\subsubsection{Control Message Channels}

In the hybrid control model for LazyCtrl, there are three types of control message channels, \emph{i.e.}, logical links.
\begin{itemize}
\item[$\triangleright$] {\it Control link.}~~A control link refers to a logical control channel (an IP tunnel or a TCP/SSH connection on top of the underlay network) via which the controller receives forwarding requests, and/or sends commands or rules to individual edge switches. The control link is extended from the secure channel between an OpenFlow controller and an OpenFlow switch by allowing the exchange of switch grouping and other related messages. When a control task cannot be handled by local control groups, packets will be forwarded to the controller and the controller will react to the edge switches by sending them flow rules or other commands, all through the control link.

\item[$\triangleright$] {\it State link.}~~A state link is a logical communication channel between the controller and a designated switch. The designated switch in each group aggregates the network states it collects from other edge switches in the group and reports them to the controller periodically via the state link. Thus, global and coherent visibility can be achieved at the controller.

\item[$\triangleright$] {\it Peer link.}~~A peer link refers to a logical control channels used for disseminating network states for address learning and updating among the switches in the same local control group. In principle, peer links would rely on multicasting. However, assuming native multicast support for the underlay may not be practical. Therefore, our design adopts an alternative approach: the designated switch (or its backup, if any) gathers network states from every peer edge switch and then disseminates them to all other switches in the same group with multiple unicast messages.
\end{itemize}

\subsection{Switch Grouping}
\label{subsec:grouping}

The design of LazyCtrl is based on the concept of grouping switches to form multiple local control groups. Thus the quality of efficiency of the grouping is essential to the whole design. Given a limit for the group size (determined according to empirical or historical data), a good grouping scheme is defined as one in which the inter-group traffic is small (in order to facilitate the laziness of the controller) and the computational complexity of the grouping algorithm is sufficiently low such that it can fast adapt to traffic dynamics. Our grouping algorithm aims at satisfying the above principles and we base our design on solving the classical graph partition problem, with improvements on time complexity and support for incremental updates.

\subsubsection{Problem Modeling}
Denote by $\mathbf{S} = \{S_1,S_2,...,S_N\}$ the set of edge switches in the multi-tenant data center network. Let $\mathbf{W} = \{w_{i,j}~|~S_i,S_j \in \mathbf{S}\}$ be an intensity matrix where each element $w_{i,j}$ represents the normalized traffic intensity (\emph{i.e.}, number of new flows per second) between two edge switches $S_i$ and $S_j$. A grouping scheme $\mathbf{G}$ is a series of disjoint subsets of edges switches, which can be defined by $\mathbf{G} = \{G_1, G_2,..., G_K~|~(G_i \subseteq \mathbf{S}) \wedge (G_i \cap G_j = \emptyset) \}$. Then, the normalized inter-group traffic intensity (denoted by $W_{\rm inter}$) can be represented by
\begin{equation}
W_{\rm inter} = \sum_{\{x,y \in [1,...,K] \wedge x \neq y\}} \sum_{\{S_m \in G_x, S_n \in G_y\}} w_{mn}. \nonumber
\end{equation}
Given an intensity matrix $\mathbf{W}$, the goal of the switch grouping problem is to find out a grouping scheme $\mathbf{G}$ such that the inter-group traffic intensity $W_{\rm inter}$ is minimized. This problem is similar to the graph partition problem where the goal is to partition a given graph into $k$ roughly equal components such that the total weigh of the edges connecting the vertices in different components is minimized (called $k$-way partitioning). The graph partition problem has been shown to be NP-hard \cite{Karypis_Kumar-1998}. The switch grouping problem differs slightly from the graph partition problem in terms of that the largest size of a group is strictly contained by a constant while the number of groups is variable.

\subsubsection{Solving the Switch Grouping Problem}

Our design for the switch grouping algorithm is based on the Multi-Level $k$-way Partition (MLkP) algorithm proposed by Karypis and Kumar for fast $k$-way partitioning for a given graph \cite{Karypis_Kumar-1998}. MLkP first reduces the size of the graph by collapsing vertices and edges. When a $k$-way partitioning of the smaller collapsed graph is found, the algorithm uncoarsens and refines this partitioning to construct a $k$-way partitioning for the original graph. The running time of MLkP is linear in the number of edges in the graph. However, direct application of MLkP to the switch grouping problem may lead to infeasible solutions, \emph{i.e.}, the sizes of the resulted partitions may exceed the given group size limit. 

We propose SGI, a Size-constrained Grouping algorithm with Incremental update support. In the initial stage (function {\tt IniGroup}), SGI first determines the right number $k$ of groups to be generated. This value can be estimated by the number switches divided by the group size limit. Next, SGI constructs an intensity graph where the vertices in the graph represent all the switches while each edge represents the communication between the two end switches of this edge. The weight on each edge indicates the normalized traffic intensity between any pair of switches, which is estimated based on history traffic statistics. Then, an initial feasible grouping of the switches is produced by using the MLkP algorithm with the constructed graph as input. Hereafter, SGI keeps running by monitoring the traffic condition on the network. Upon a significant change\footnote{The controller evaluates the significance of traffic change by measuring the difference in its workloads.} on the traffic distribution, SGI carries out a greedy refinement function called {\tt IncUpdate} to incrementally update the grouping in order to reduce the inter-group traffic. The refinement process runs iteratively and in each iteration, two groups between which traffic volume increases the most are merged and split again to ensure minimized communication between the two new groups. This is identical to finding a minimum bisection cut of a given graph, which can be accomplished efficiently in polynomial time \cite{Stoer_Wagner-1997}. The refinement process will terminate when the workload of the controller meets some threshold. The pseudocode of the SGI algorithm is given in Fig.~\ref{fig:sgi}.

\begin{figure}[!t]
{\tt \scriptsize
\begin{verbatim}
IniGroup:
 1: // construct the intensity graph
 2: ConstructGraph(history intensity matrix)
 3: // obtain the initial grouping
 4: MLkP(intensity graph, #partition k)

IncUpdate:
 5: // running in background
 6: while(true):
 7:   // the controller is overloaded
 8:   while (controller.load > threshold.high):
 9:     // find two candidate groups with
10:     // the most significant traffic change
11:     FindGroups(all groups)
12:     MergeGroups(candidate groups)
13:     SplitGroup(the combined group)
14:     // the controller is underloaded
15:     if (controller.load < threshold.low):
16:       break
\end{verbatim}
}

\caption{\label{fig:sgi}Pseudocode for the SGI algorithm.}
\vspace{-0.5cm}

\end{figure}

\subsection{Packet Forwarding}

\subsubsection{Setup Phase}

Similar to that of typical OpenFlow control, in LazyCtrl, the edge switches are configured to point to the central controller at the setup phase. Besides generating the local control groups by invoking the SGI algorithm, the controller is also in charge of the following configurations for every group before the whole LazyCtrl system comes into function.

{\it Selecting designated switches.}~~For each local control group, the controller selects a designated switch among all edge switches in this group by applying some given principle such as shortest physical distance, shortest response time to the controller. If necessary, the selection process also includes choosing some backups for the designated switches.

{\it Ordering and informing edge switches.}~~The controller orders all switches in a group according to the physical (MAC) address of switch's management interface. This is for building a logical ring for failure auto-detection (detailed in Section \ref{subsec:failover}). The controller then delivers to each switch its neighbors on the logical ring. Besides that, the controller will also inform the switches in a group with the designated switch ID and some global timing and performance parameters such as the group size, the frequency to apply group synchronization or keep-alive heartbeats.

\subsubsection{Table Organization}

The core--edge separation enables one-hop ``logical" distance between any pair of edge switches, leaving basic packet routing to the IP underlay.  What remains unsolved is the host-to-switch mapping.

\begin{figure}[!t]
\centering
\includegraphics[scale=0.35]{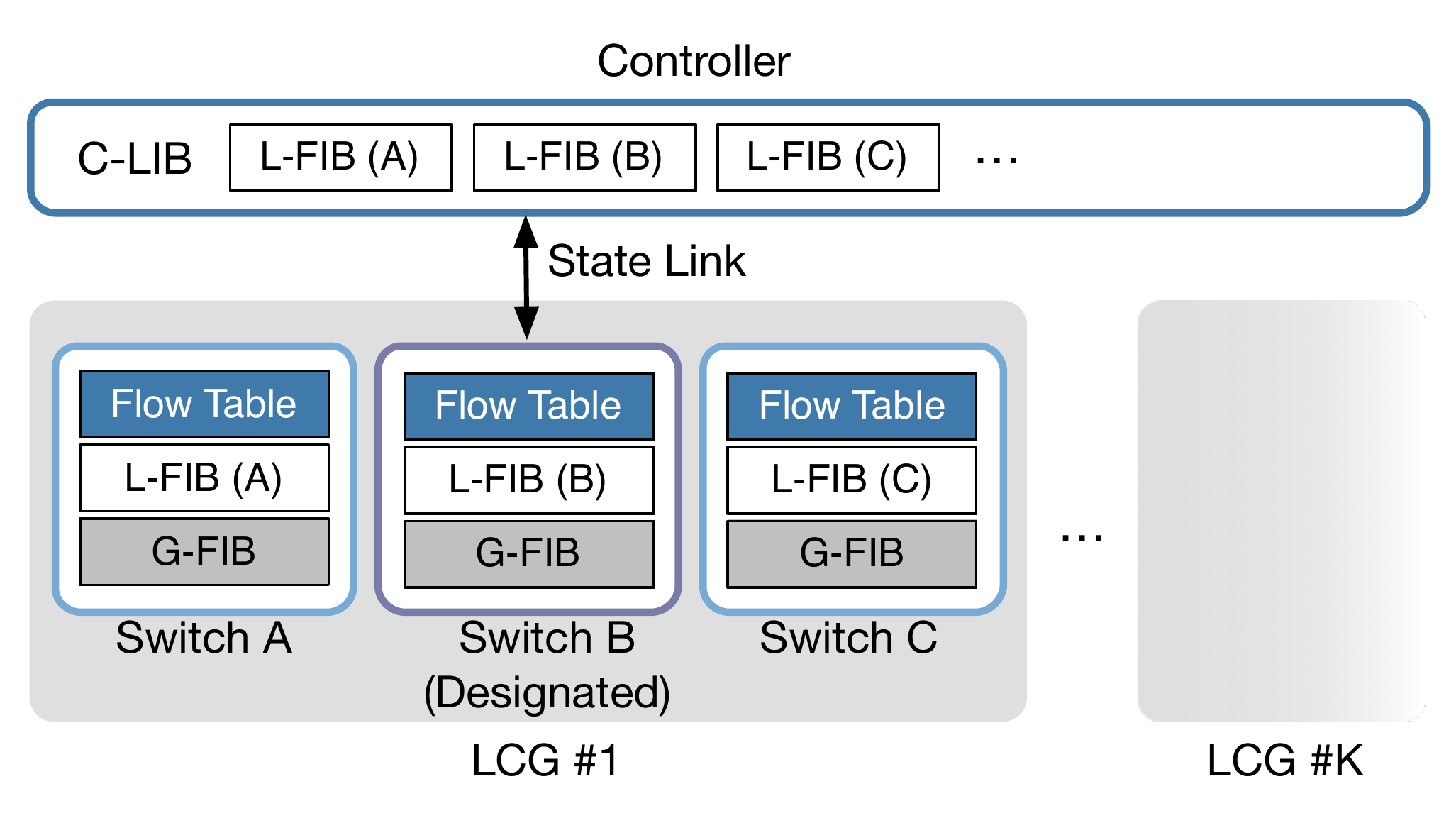}

\caption{\label{fig:table} Table organization in the LazyCtrl design.}
\vspace{-0.5cm}

\end{figure}

In the LazyCtrl design, each edge switch is associated with a Local Forwarding Information Base (L-FIB), which tracks the hosts or virtual machines that are attached to this switch. To handle intra-group traffic, each edge switch also maintains a replica of the L-FIBs of all other switches in the same group, which we call Group Forwarding Information Base (G-FIB). The central controller retains global visibility of the network by maintaining a Central Location Information Base (C-LIB), which contains the L-FIBs of all edge switches in the network. Using this C-LIB, the controller can handle inter-group traffic and any other specific flows whose control requires global visibility. A general overview of the table organization in the LazyCtrl design is depicted in Fig.~\ref{fig:table}.

{\bf L-FIB:} The L-FIB of each edge switch is implemented with a conventional lookup mechanism similar to the MAC/ARP table in ordinary layer two switches. 

{\bf G-FIB:} The G-FIB of each edge switch is a replica of the L-FIBs of all switches in the same group. To save storage space, we implement G-FIB using Bloom Filter (BF), as the storage space required by a BF is independent from the number of elements it contains. The G-FIB of each edge switch is comprised of multiple BFs generated from the L-FIBs of all switches in this group. Given an address of a virtual machine, each BF decides whether this address is under the corresponding edge switch. All the BFs together will return a vector of Boolean values indicating the possible location of this address. Note that it might happen that there are multiple possible locations for one address, which is resulted from the \emph{false positive} of BFs. However, the false positive rate is predictable and controllable by space-time trade-offs \cite{Bloom-1970}.

\subsubsection{State Dissemination}

State dissemination consists of the mechanisms to spread and synchronize network states (\emph{e.g.}, the host-to-switch mapping) and updates in the control plane. In general, there are two types of state dissemination in LazyCtrl:

{\it Live/Synchronized state dissemination.}~~Live state dissemination refers to the host discovery process driven by the end hosts via ARP broadcasting in the bootstrapping stage and at virtual machine migration or removal. In the LazyCtrl design, live state dissemination can be cascaded in three different levels: $\mathit{i})$ Upon receiving an ARP request, the edge switch learns the source address by inserting or updating an item in its L-FIB and then floods the request to all relevant local ports. $\mathit{ii})$ If no local hosts (that are attached to this edge switch) answer this request and the requested destination cannot be recognized by the G-FIB of the switch either, this request will be forwarded to the designated switch in this group for an intra-group ``broadcasting''. $\mathit{iii})$ Further, if there is still no response from the hosts in this group, the request will be forwarded to the central controller, which relays the request to the designated switches in all other groups that contain hosts belonging to the relevant tenant (\emph{e.g.}, according to tenants' VLAN settings).

{\it Asynchronous state dissemination.}~~When the traffic pattern changes, the grouping of the switches may not be effective for shielding the central controller thus needs to be adjusted. The condition of inter- and intra-group traffic is also changed. Therefore, the host-to-switch mapping must be re-disseminated across the control plane in order for local control groups to handle all intra-group traffic. This is different from the case of virtual machine migration in the sense that there is no change to the host-to-switch mapping. As a result, the end hosts cannot sense this change and thus cannot accordingly drive any updates. Moreover, in an extreme case where all hosts from a certain tenant appear in the same local control group, the controller may want to block all ARP request from that tenant to avoid unnecessary workload of itself. However, this could lead to incomplete visibility for the controller as the traffic from that tenant will be transparent to the controller.

In order to handle the above circumstances, an asynchronized switch-driven state dissemination mechanism must be introduced. In LazyCtrl this mechanism contains two aspects: $\mathit{i})$ When an update event occurs at an edge switch, this switch sends its updated L-FIB to the designated switch in the group via the peer link; the designated switch then relays this update to all the other switches in the same group to synchronize the group-wide network state. The designated switch then sends the update to the controller via the state link to synchronize the network state between the controller and the local control group. $\mathit{ii})$ When the grouping of the switches has been changed, the controller sends the L-FIBs of the switches in a new group to the designated switches in this group via the state links. The designated switch then ``broadcast'' the L-FIBs to all the edge switches in this group for updating their G-FIBs.

\subsubsection{Packet Forwarding Routine}

\begin{figure}[!t]
{\tt \scriptsize
\begin{verbatim}
Upon arrival of a packet P at an edge switch S:
 1: // P originates from a local host
 2: if (P is a local plain packet):
 3:   // handled by flow table
 4:   if (P matches S.flow_table.rule):
 5:     apply S.flow_table.action to P
 6:   // handled by local control group
 7:   else:
 8:     // lookup in the L-FIB
 9:     host = LookUp(L-FIB, P.dest_addr)
10:     // no match found
11:     if (host == none):
12:       // query in the G-FIB
13:       dst_vec = Query(G-FIB, P.dst_addr)
14:       // no match found, handled by controller
15:       if (dst_vec == empty):
16:         send P to the controller
17:       // send P to all possible targets
18:       else: for each (dst in dst_vec):
19:         encap. and send a copy of P to dst
20:     // match a local host
21:     else: forward P to host
22: // P is an encapsulated packet
23: else:
24:   Decapsulate(P)
25:   // lookup in the L-FIB
26:   host = LookUp(L-FIB, P.dst_addr)
27:   // no match found (due to false positive)
28:   if (host == none): Drop(P)
29:   else: forward P to host
\end{verbatim}
}
\vspace{-0.3cm}
\caption{\label{fig:fwd}Packet forwarding routine in LazyCtrl.}
\vspace{-0.5cm}

\end{figure}

We describe now how traffic control is carried out in LazyCtrl. The detailed forwarding routine of a packet is shown in Fig.~\ref{fig:fwd}. When a packet arrives at an edge switch, depending on packet type, the following two actions will be applied: 
$\mathit{i})$ If the packet is plain (which originates from a local host), the switch first carries out a lookup in its flow table to check whether there are matched rules for this packet. If so, the action corresponding to the rule is then applied to the packet; otherwise, the switch continues looking up in its L-FIB to check whether the destination of this packet is a local host. A packet with an address of a local host will be forwarded directly to that host. If no entry matched the L-FIB, the switch carries out a query in its G-FIB. Note that there might be multiple targets for this packet returned from this query due to the false positive of BFs. The switches then send to all the targets a copy of the packet. If all the elements in the Boolean vector are false, it means that the target of this packet is not in the current group and thus the packet will be forwarded to the controller to request inter-group control rules. $\mathit{ii})$ If the packet is encapsulated, the switch first decapsulates it and then carries out a lookup in its L-FIB to determine its destination host. If no matched entries are found, the switch simply drops the packet as it knows that this packet is mis-forwarded to the switch due to BF's false positive. Optionally, this mis-forwarded packet could also be directed to the controller for installing flow entries on related switches to avoid further false positive for the same destination.

\subsection{Failover}
\label{subsec:failover}

\subsubsection{Failure Detection}

The switch grouping scheme ensures that switches in the same group are ``strongly connected'' due to their frequent traffic exchange. As a result, failures in the data plane can be passively detected quickly. In contrast, handling failures in the control plane is more laborious.

In the LazyCtrl design, we propose a self-detection mechanism to handle failures in the control plane based on a group-wide failure-detection wheel with the controller at the center and the switches at the edge. As we have mentioned previously, at the setup phase the controller orders the switches to form a wheel and informing the switches in the same group their neighbors on the wheel. To detect failures, keep-alive messages will be initiated from upstream switches to downstream switches and from the controller to each switch. All possible cases of failures depending on the place of packet loss are listed in Table~\ref{tb:failure}.

\begin{table}[!t]
\centering
\scriptsize
\begin{tabular}{c|c|c|c}
\hline
\multirow{2}{*}{\bf Failure} 
& \multicolumn{3}{c}{\bf Packet loss}  \\ \cline{2-4}
& $S_n \rightarrow S_{n-1}$ & $S_n \rightarrow S_{n+1}$ & Controller$\rightarrow S_n$ \\ \hline\hline
Control link & ~ & ~ & $\checkmark$ \\
Peer link (Up) & $\checkmark$ & ~ \\
Peer link (Down) & ~ & $\checkmark$ & ~ \\
Switch ($S_n$) & $\checkmark$ & $\checkmark$ & $\checkmark$ \\
\hline
\end{tabular}
\caption{\label{tb:failure} Inferring failures in the control plane according to the place of packet loss.}
\label{table2}
\vspace{-0.9cm}
\end{table}

\subsubsection{Failover of Links}

Link failures indicate routing-related issues, \emph{e.g.}, packet loss due to link congestion or temporary routing loops on the underlay. We adopt detour routing based approaches to handle link failures in LazyCtrl. When a data path failure occurs, for instance, between $S_n$ and $S_{n-1}$, the controller will be notified and an alternative path will be chosen for delivering packets following $S_{n} \rightarrow S_{n-1}$. For a failure on the control link between the controller and a switch such as $S_n$, the controller will send a request to the upstream switch of $S_n$ on the failure-detection wheel, \emph{i.e.}, $S_{n-1}$, to pass on the control message from $S_n$ to the controller. When a peer link failure occurs (between $S_n$ and $S_{n-1}$), the control functionality is affected only when one of the two end switches is the designated switch. In this case, the controller will ask $S_n$ or $S_{n-1}$ to quit as the designated switch and reselects one from the backups for the designated switch to fulfill the role of designated switch.

\subsubsection{Failover of Switches}

A switch failure usually turns to be a reboot or a reset of the switch, especially in the case where edge switches are implemented with virtual switches in hypervisors. The controller is responsible for detecting the malfunction of the switch and then carries out the following actions: $\mathit{i})$ informing the designated switch in the same group this switch failure and asking the designated switch to spread the temporary outage of the failed switch in the group in order to avoid unexpected detour routing requests; $\mathit{ii})$ rebooting the failed switch remotely and checking its comeback periodically; $\mathit{iii})$ removing the outage signal and proactively triggering a state synchronization in the group when the switch is back to function.

If the failed switch is the designated switch in the group, in addition to the above actions, the controller will select a new designated switch for the group. If backups are set for the designated switch, no single point of failure exists since those backups work simultaneously and will be fixed upon a failure independently.

\section{Implementation}
\label{sec:implementation}

We implement LazyCtrl by extending the OpenFlow protocol and developing edge switches and the controller based on Open vSwitch \cite{OVS} and Floodlight \cite{Floodlight}. The source code of our implementation can be found on \cite{LazyCtrl}.

\subsection{Open vSwitch-based Edge Switch}
The main forwarding component of Open vSwitch consists of the {\tt ovs-vswitchd} and datapath modules. The {\tt ovs-vswitchd} module works in the user space, handling slow path processing such as learning, remote configuration, full flow-table lookup; the datapath module in the kernel space handles fast path processing including packet forwarding, quick-table lookup, modification, and tunneling. The implementation of the LazyCtrl edge switch follows a similar design principle. Fast path processing, such as L-FIB lookup (including BF matching), packet encapsulation, and forwarding, are integrated into the kernel space (datapath) module while a few slow path modules are integrated into {\tt ovs-vswitchd} which are listed as follows.

\begin{itemize}
\item[$\triangleright$] {\it Ctrl-IF module} is an interface for the switch to interact with the controller, which also implements the control link. Unknown packets (from inter-group traffic) will be forwarded to the controller using OpenFlow {\tt Packet\_In} messages.
\item[$\triangleright$] {\it State advertisement module} is introduced for collecting and disseminating local host information and traffic statistics among the switches in the same group.
\item[$\triangleright$] {\it FIB maintenance module} maintains the L-FIB and the Bloom filter based G-FIB structures according to the network states collected by the state advertisement module and then updates the kernel space module for fast path processing.
\item[$\triangleright$] {\it State reporting module} will only be activated when the switch is selected as the designated switch for the group. This module implements all functions associated with the state link.
\end{itemize}

\subsection{Floodlight-based Controller}

The Floodlight OpenFlow controller provides a rich set of components. The central controller in LazyCtrl is implemented based on the existing Floodlight controller by introducing the following extensions.

\begin{itemize}
\item[$\triangleright$] {\it Encap action} realizes packet encapsulation in edge switches by extending the existing OpenFlow v1.0 protocol. In the LazyCtrl architecture, packet forwarding in the data plane overlay replies on a GRE-like encapsulation. When a rule with this action is applied to a flow, the switch will encapsulate the packets with a new header targeting a given remote IP address.
\item[$\triangleright$] {\it C-LIB maintenance module} implements the functions of acquiring L-FIBs from the designated switch in every group and building the C-LIB at the controller.
\item[$\triangleright$] {\it Switch grouping management module} handles the management of the local control groups. We base our implementation of switch grouping on the proposed SGI algorithm. A daemon module is introduced to handle the state reports from the designated switches in all groups and keep analyzing the changes in traffic pattern. Re-grouping will be triggered when $\mathit{i})$ the workload of the controller suffers from an accumulated growth of up to $30\%$ from last update or $\mathit{ii})$ it has been two minutes since last update. Setting up a minimum update interval (2 minutes here) is to prevent the oscillation caused by short-term traffic fluctuation.
\item[$\triangleright$] {\it Tenant information management module} is used to manage tenant information such as VLAN IDs in switches. Being aware of this information, the controller can determine where to spread the ARP messages and when inter-group traffic control is necessary.
\item[$\triangleright$] {\it Failover module} is in charge of failure detection and recovery as we have discussed in Section~\ref{subsec:failover}.
\end{itemize}

\section{Evaluation}
\label{sec:evaluation}

\subsection{Prototype Setup}

To validate the performance of the LazyCtrl implementation, we conducted experiments based on a real traffic trace collected from an enterprise production data center in Europe, which consists of 272 GigE edge switches and 6509 hosts. Accordingly, we built a prototype system using 6 Pronto 3290 switches and 24 IBM x3550 8-core (two quad cores) servers. The switches were interconnected with a full mesh via 10 GigE links, severing as the network core (IP-based underlay). Each switch was connected with 4 servers via GigE links. To emulate the 272 edge switches in the real data center, we deployed 272 Linux virtual hosts running our modified Open vSwitch implementation on the 24 servers. A custom-made trace re-player was developed and deployed on each of the 272 Linux virtual host to replay the inter-switch traffic generated by the 6509 hosts in the trace. The Floodlight-based central controller was hosted on a standalone Linux PC (with Intel Core 2 Duo CPU 2.2 GHz) and could be configured to run in either lazy or normal mode.

\subsection{Datasets}

\begin{table}[!t]
\centering
\vspace{0.3cm}
\small
\begin{tabular}{c| c| c| c| c}
\hline
{\bf Trace} & {\bf \# of flows} & {\bf Avg. centrality} & $\mathbf{p~(\%)}$ & $\mathbf{q~(\%)}$ \\ \hline\hline
Real & 271M & 0.85 & N/A & N/A \\ 
Syn-A &2720M & 0.85 & 90 & 10 \\ 
Syn-B & 3806M & 0.72 & 70 &20 \\ 
Syn-C & 5071M & 0.61 & 70 & 30 \\
\hline

\end{tabular}
\caption{\label{tb:trace} Characteristics of the traffic traces.}
\vspace{-0.9cm}
\end{table}

The real traffic trace we collected consists of the traffic among 272 GigE edge switches and 6509 hosts over a whole day. To check the consistency of the performance results under different traffic scenarios, we generated three synthetic traffic traces based on the real trace. The main characteristics of all traces are summarized in Table~\ref{tb:trace}. In the synthetic traces, traffic is assumed to be exchanged through 2713 edge switches among 65090 hosts, with a scaling-up factor of 10 compared with the real trace. The traffic flows in the synthetic traces were generated in the following manner so that the key characteristics such as the temporal patterns could be retained: $p\%$ of the flows are generated by selecting from a given set of host pairs ($q\%$ of all host pairs) uniformly at random in the synthetic topology, and assigning each selected host pair a payload randomly chosen from the real trace. We vary the values for $p$ and $q$ and three synthetic traces are generated with significant differences in traffic locality represented by average centrality. The rest flows are generated by selecting host pairs uniformly at random from all host paris in the synthetic topology. Each selected host pair is assigned with a payload randomly chosen from the real trace.

\subsection{Performance of Switch Grouping}

\begin{figure}
	\centering
	\subfigure[]{
 	\label{fig:grouping_quality} 
	\includegraphics[scale=0.23]{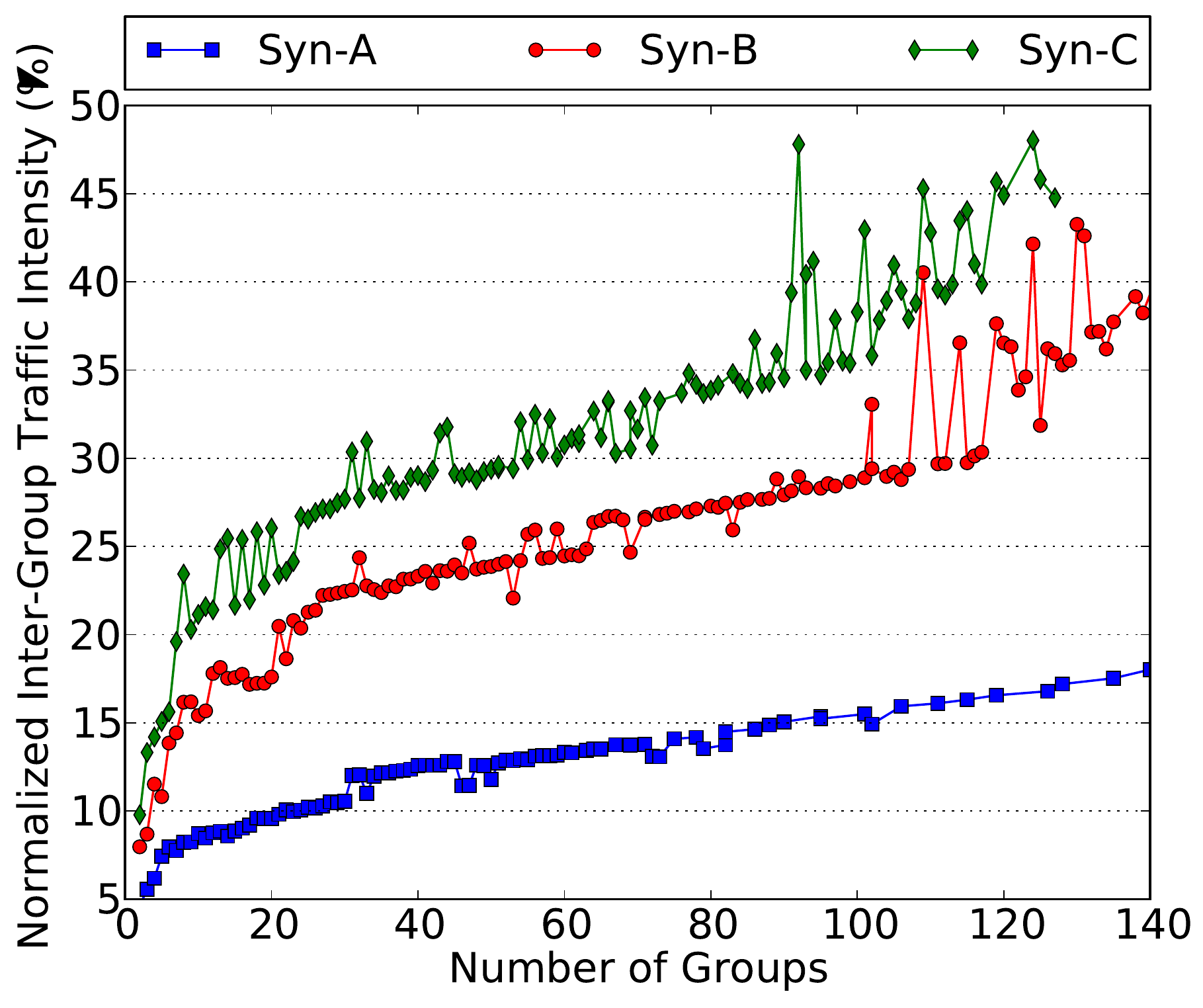}}
	\hspace{-0.1in}
	\subfigure[]{
	\label{fig:grouping_time} 
	\includegraphics[scale=0.23]{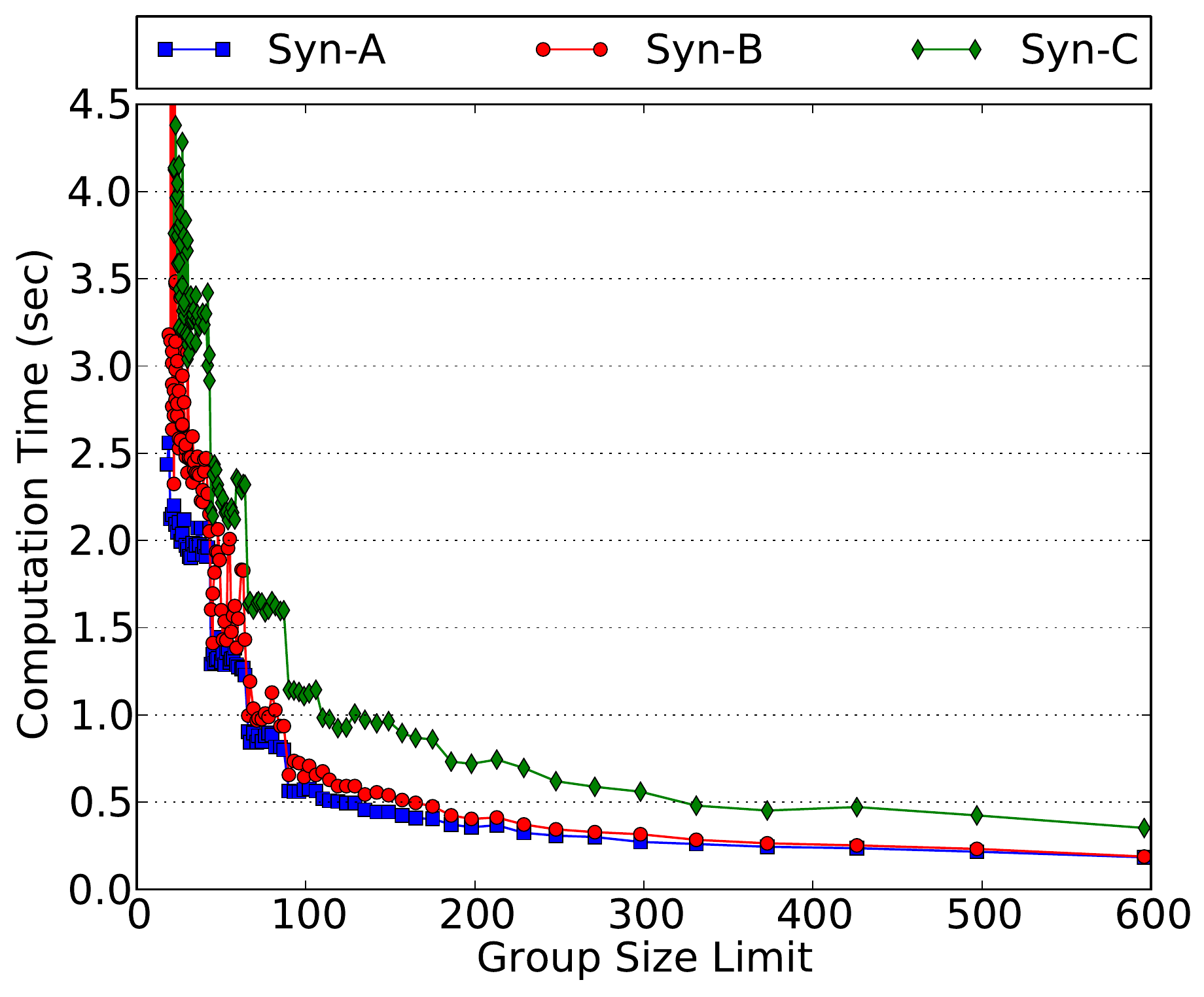}}
	\vspace{-0.4cm}
	\caption{a) Normalized inter-group traffic intensity when grouping with different numbers of groups on each of the three synthetic traffic traces. b) Computation time of switch grouping under different group size limits on each of the three synthetic traffic traces.}
\vspace{-0.5cm}

\end{figure}

We evaluate the quality of the proposed switch grouping scheme by calculating the normalized inter-group traffic intensity ($W_{\rm inter}$ as defined in Section~\ref{subsec:grouping}). Fig.~\ref{fig:grouping_quality} depicts the results of applying the size-constrained MLkP algorithm (the {\tt IniGroup} function in SGI) to the traffic derived from each of the three synthetic traces with various numbers of groups. We observe that the grouping quality varies across different traces. In general for traces with higher average centralities, it tends to have smaller values for $W_{\rm inter}$, indicating better performance in reducing the workload of the controller. We also observe that $W_{\rm inter}$ increases almost linearly with the increase of the number of groups, confirming that maximizing the sizes of single groups (thus consequently reducing the number of groups) will best facilitate the laziness of the controller.

\begin{figure*} [!t]
  \begin{minipage}[t]{0.31\linewidth} 
    \centering 
    \includegraphics[scale=0.24]{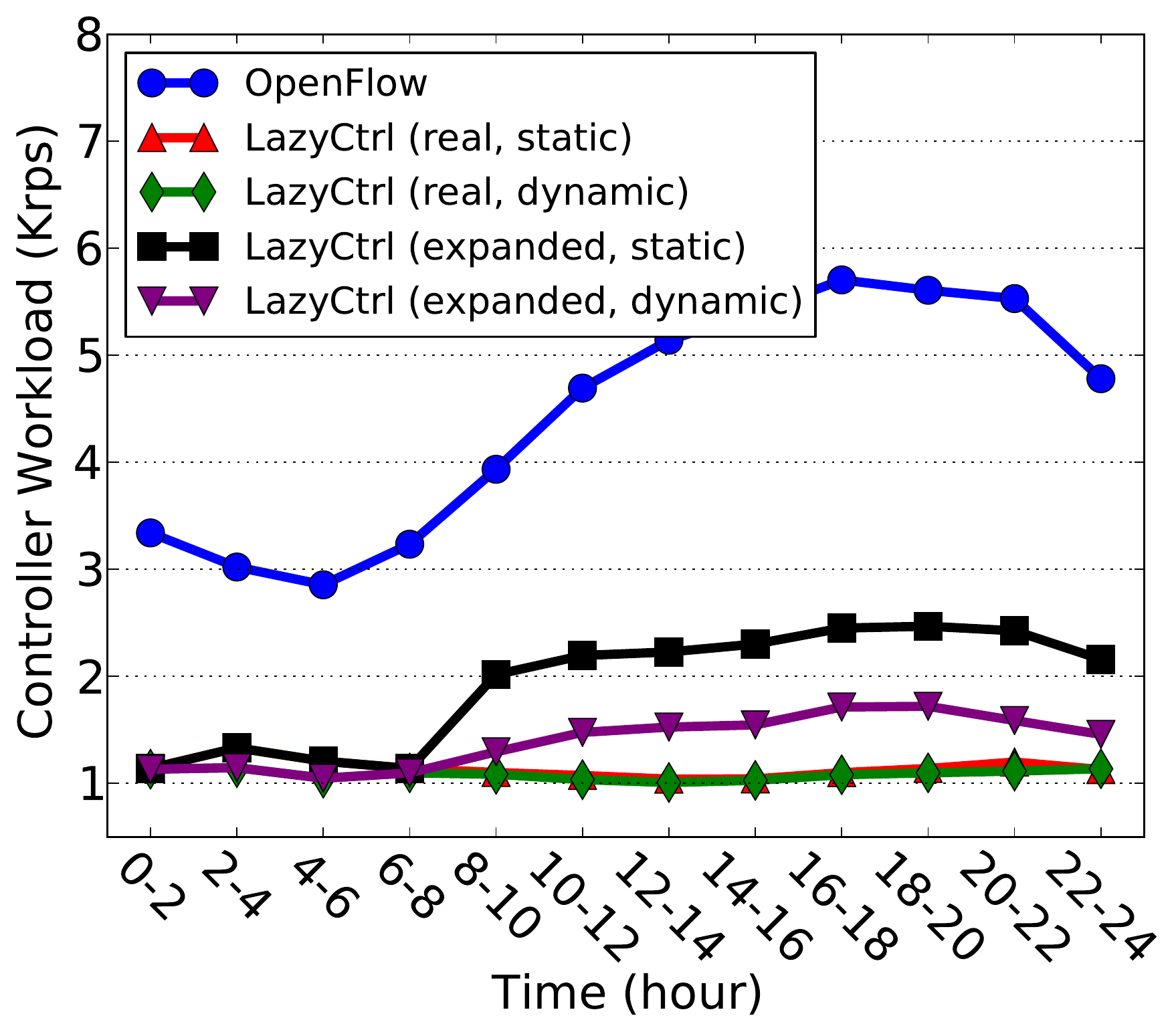} 
    \caption{Controller workload.} 
    \label{fig:workload} 
  \end{minipage}%
  \hspace{0.3cm}
  \begin{minipage}[t]{0.31\linewidth} 
    \centering 
    \includegraphics[scale=0.24]{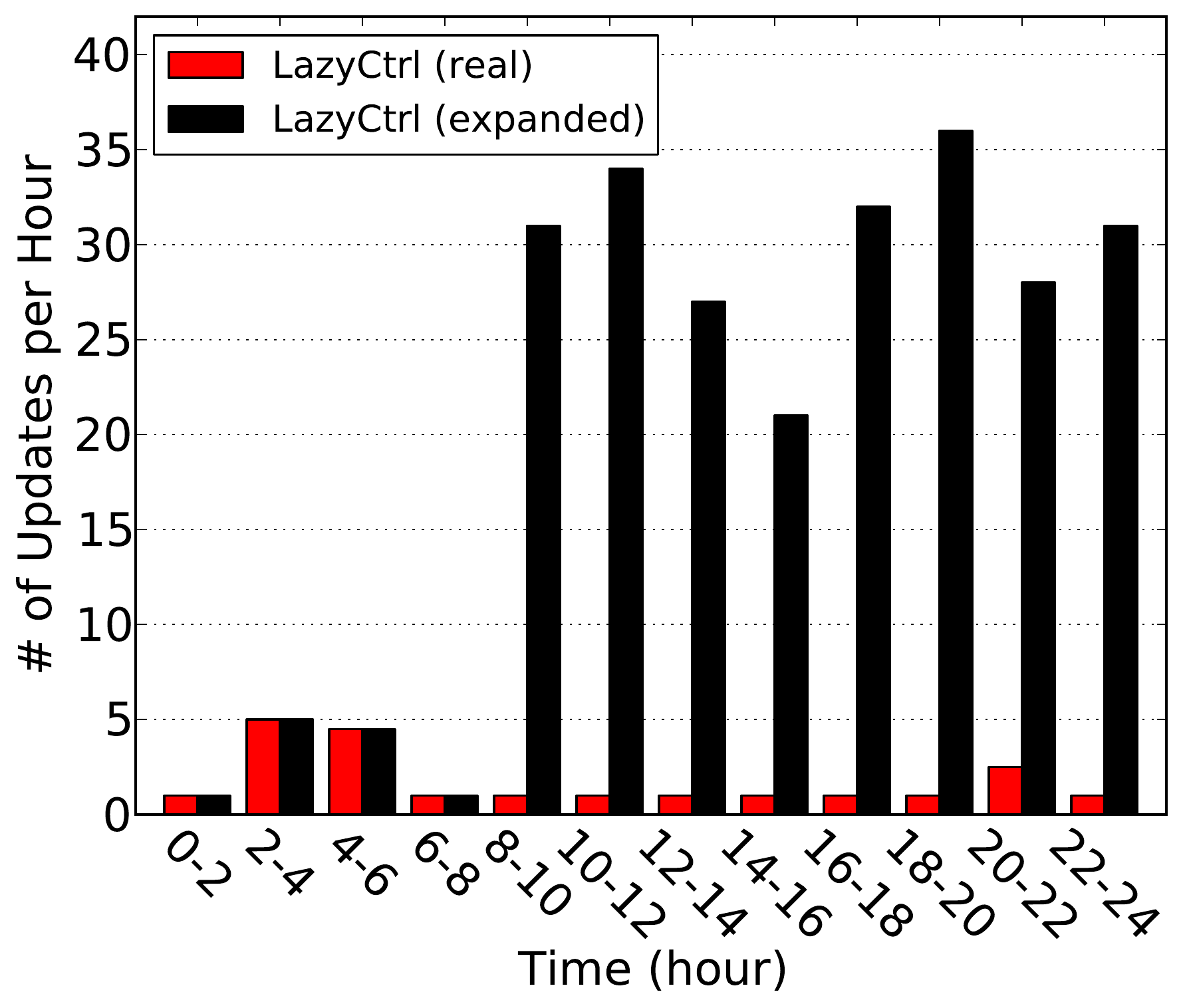} 
    \caption{Switch grouping frequency.} 
    \label{fig:frequency} 
  \end{minipage} 
  \hspace{0.3cm}
  \begin{minipage}[t]{0.31\linewidth} 
    \centering 
    \includegraphics[scale=0.24]{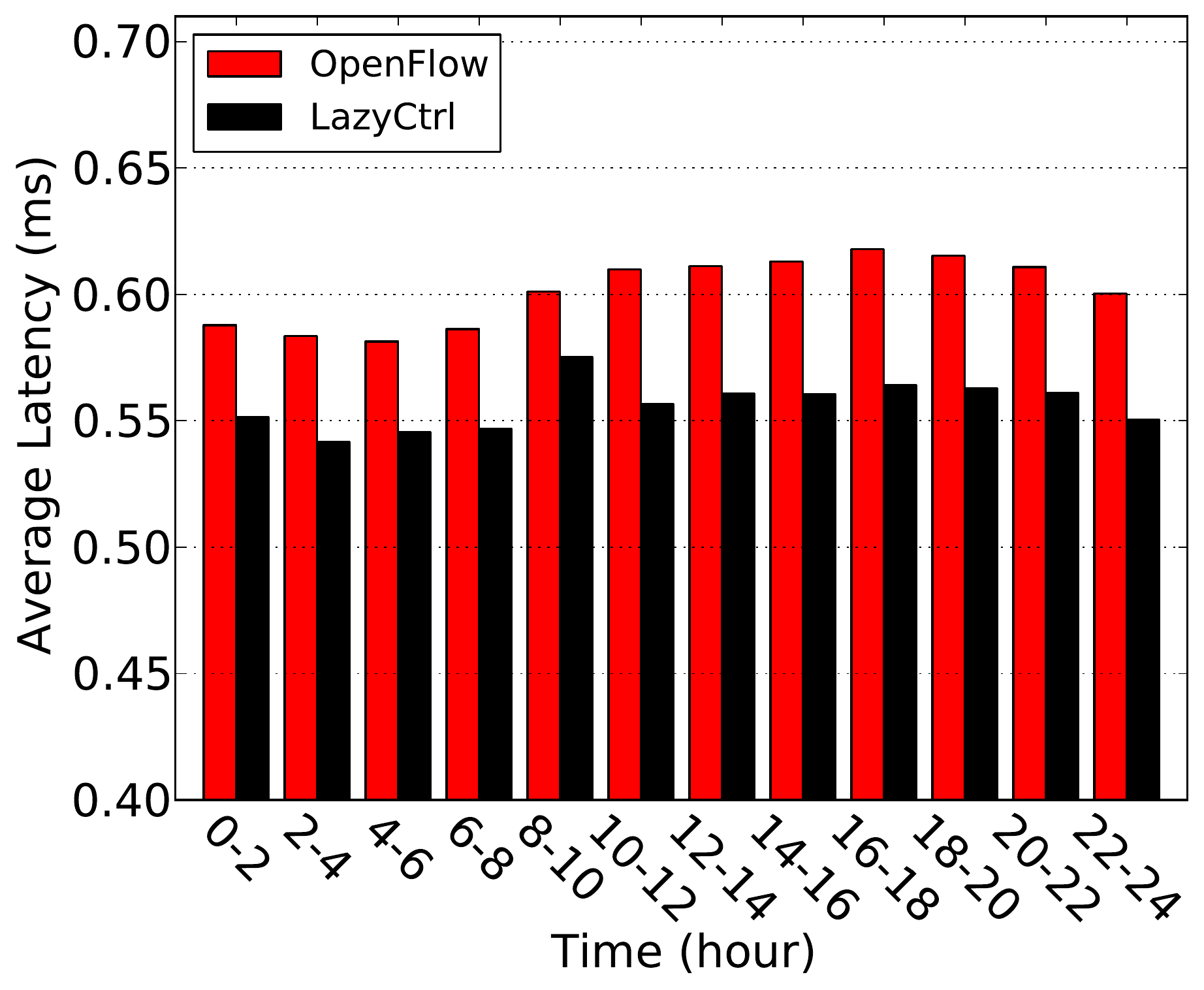} 
    
    \caption{Steady state latency.} 
    \label{fig:latency} 
  \end{minipage}
  \vspace{-0.5cm}
\end{figure*}

We also carry out measurements to examine the computation time in switch grouping generation on the three synthetic traffic traces. The results of applying the {\tt IniGroup} function in SGI with various group size limits are shown in Fig.~\ref{fig:grouping_time}. It can be seen that switch grouping can be accomplished as fast as less than 5 seconds and the grouping time is inversely proportional to group size limit. Note that switch grouping is only carried out when there is a very significant change in traffic pattern and incremental updates are not possible to retain good grouping quality in reasonable time. During most of the time, applying the {\tt IncUpdate} function is sufficient, which is more than one order of magnitude faster than the {\tt IniGroup} function.

\subsection{Effectiveness of LazyCtrl}

{\it Controller workload.}~~We validate the effectiveness of LazyCtrl by measuring the controller workload under traffic dynamics. We first conduct a comparison to standard OpenFlow control (with the original Floodlight implementation) using the real traffic trace. For LazyCtrl, the initial grouping is done based on the first-hour traffic pattern and we test in both \emph{static} and \emph{dynamic} cases with and without incremental updates for the grouping, respectively. To further verify the consistency of the results, we expand the real trace by introducing $30\%$ extra flows among the hosts that did not communicate with each other in the real trace during the time interval from 8 to 24. Using the \emph{expanded} trace, we test again in both static and dynamic cases. We compare the workload of the controller in the above cases and the experimental results are illustrated in Fig.~\ref{fig:workload}. It can be observed that $\mathit{i})$ LazyCtrl can help achieve a significant level of workload reduction (about $61\%$--$82\%$) for the controller; $\mathit{ii})$ The controller workload in LazyCtrl is relatively stable during the day on the real trace, which is due to the fact that majority of the traffic growth happens among those ``strongly connected'' hosts inside local control groups, being transparent to the controller. $\mathit{iii})$ The controller workload can be significantly reduced when the {\tt IncUpdate} function is applied due to the fact that the additionally introduced flows keep breaking the skewness of the traffic over time and thus grouping updates have to be applied continuously to adapt to the changes in order to prevent the controller from being overloaded.

{\it Grouping update.}~~In addition, we examine the update frequency of switch grouping on both the real and expanded traces. The update frequency results are shown in Fig.~\ref{fig:frequency}. It can be noticed that the incremental update function has very limited influence on the controller workload on the real trace. At the same time, the update frequency keeps at a very low level (10 updates per hour), indicating that maintaining a relatively effective grouping is feasible in practice. On the expanded trace, the cost for keeping the controller lazy is a reasonable increase in update frequency (with a maximum of 34 updates per hour).

{\it Storage overhead.}~~The storage cost of the BF-based G-FIB on each switch is linear with the group size. For example, when a group consists of 46 switches, for each switch the BF-based G-FIB contains 45 bloom filters. Assuming that each bloom filter has $16$ $128$-byte entries, the memory required for the BF-based L-FIB on each switch is $45\times16\times128 = 92,160$ bytes, resulting in a false positive rate of less than $0.1\%$.

\subsection{Latency Overhead}

{\it Cold-cache forwarding latency.}~~We evaluate the forwarding latency under ``cold-cache'' scenarios upon the first packet of a fresh flow is injected into the network. We emulate cold-cache scenarios by launching 45 new flows among 5 newly deployed hosts and compare the average forwarding latency of the first packets of these flows in LazyCtrl to that in the standard OpenFlow control. For intra-group traffic, the cold-cache forwarding latency in LazyCtrl (0.83 ms) is more than an order of magnitude smaller than that in OpenFlow (15.06 ms). This is due to the fact that packets from intra-group traffic will be forwarded locally without involving the controller. The data plane operations such as L-FIB lookup and packet encapsulation are very fast and thus packets can be processed at line speed. For inter-group traffic, LazyCtrl also outperforms standard OpenFlow by achieving a cold-cache latency of 5.38 ms. This is because LazyCtrl requires no passive learning of the network topology through all ARP flooding as is the case of standard OpenFlow (the {\tt learning-switch} module in Floodlight), which is another benefit brought by the lazy principle in LazyCtrl. 

{\it Steady-state latency.}~~Steady-state latency refers to the average forwarding latency of all processed packets over a relatively long period of time (2 hours here). The experimental results on the real trace with a $24$-hour span are illustrated in Fig.~\ref{fig:latency}. It can be observed that on average a $10\%$ reduction on latency can be achieved by LazyCtrl compared with standard OpenFlow. Moreover, this improvement is a byproduct of reducing the workload of the controller as less load on the controller leads to higher processing speed. Moreover, the synchronized state dissemination speeds up topology learning, which implicitly help reduce the response time of the controller.

\section{Conclusions}
\label{sec:conclusion}

In this paper we present LazyCtrl, a novel hybrid control plane design for data center networks. LazyCtrl is based on a core-edge separated architecture and the control functionality is implemented in a hybrid fashion: frequent coarse-grained control tasks are largely devolved to network edge by clustering edge switches into local control groups according to traffic locality and carrying out distributed control independently inside each group; the central controller is only in charge of very limited number of inter-group or other fine-grained control events. The central controller keeps adapting the grouping of edge switches to maintain its laziness. Our evaluation on the LazyCtrl prototype with both real and synthetic traffic traces show that LazyCtrl can help reduce the workload of the central controller by up to 82\%, improving the scalability of standard OpenFlow to a large extend. Moreover, LazyCtrl is fully compatible with existing solutions for scaling flow-based centralized control to large networks.






%
%
%

\bibliographystyle{IEEEtran}
\bibliography{IEEEabrv,ref}

\appendix


\subsection{Extended Discussion on Scalability}

Both distributed and centralized control approaches have some instinct limitations in scaling to large-scale data center networks. In general, distributed control such as link-state protocols depends on broadcasting to synchronize network states, which will be a disaster when the network becomes inconceivably large. Moreover, the edge switches need to learn the locations of all hosts, leading to explosive forwarding table sizes. While eliminating the need for state synchronization in principle, centralized control suffers from the stress of handling frequency and resource-exhaustive events such as flow arrivals and network-wide statistics collection events at the controller, which consequently limits the scalability of network control in data centers. 

LazyCtrl aims at finding out the right balance between distributed and centralized control and integrates the advantages from both sides to fundamentally solve the scalability issue of network control in data centers. The scalability of LazyCtrl is interpreted in three aspects:
\begin{itemize}
\item[$\triangleright$] {\it Table organization and state dissemination.}~~Due to the group size limit, the L-FIB and G-FIB on each switch will be constrained to limited sizes and thus do not have any scalability issue. Switch-wide state dissemination is also constrained in specific group and is decoupled from the size growth of the data center. The controller is designed to passively receive network states from local control groups for state dissemination and address leaching, which scales as well.

\item[$\triangleright$] {\it Location resolution.}~~Location resolution responsibility in LazyCtrl is shared among switches and the controller. Switches handle the majority of tasks locally in local control groups while the controller is only involved when intra-group control is not sufficient. As a result, the workload of the controller can be kept at a very low level, mitigating the scalability issue.
\item[$\triangleright$] {\it Failure detection and failover.}~~Clustering switches into ``strongly connected'' groups with limited sizes simplifies the process of failure detection and recovery of a large network system as failover tasks can be carried out independently in each of the groups. In addition, control authority in LazyCtrl is shared by local control groups and the controller, avoiding a single performance bottleneck.
\end{itemize}

\subsection{Possible Optimizations on Switch Grouping}

For simplicity of exposition, we omitted some optimization efforts we carried out for switch grouping in Section~\ref{subsec:grouping}. Now we highlight some of them.

\begin{itemize}
\item[$\triangleright$] {\it Host exclusion in switch grouping.}~~When a edge switch is connected to hosts belonging to many tenants, it may be difficult for a greedy method to generate a grouping with superb quality. In this case, the controller can choose some hosts and exclude them from the grouping process. The control tasks for these hosts will be accordingly handled by the controller.
\item[$\triangleright$] {\it Preload for seamless grouping update.}~~During grouping updates, the L-FIBs on the related switches will be modified, leading to forwarding interruptions. To relieve this, the controller can preload some rules to the related switches to temporarily handle the control tasks for them. These rules will be removed when the grouping becomes stable.
\item[$\triangleright$] {\it Acceleration by parallelism.}~~The {\tt IncUpdate} function in the SGI algorithm can be easily parallelized by carrying out merge and split operations simultaneously for multiple group pairs. Consequently, the computation overhead brought by the regrouping process can be further reduced.
\end{itemize}

\subsection{Methods for Determining the Right Group Size}

Determining the right sizes for groups plays an important role in keeping LazyCtrl effective. Intuitively, the larger the group size, the lower the expected workload for the controller due to less inter-group traffic. On the other hand, the larger the group size, the higher the control overhead on the switch side, as a larger group means more network states to spread among the switches in the group and more L-FIBs and G-FIBs to maintain.

Compared with empirically driven or static group sizes, we believe that a dynamic group size negotiation between the controller and the switches can be helpful, as networks can be heterogeneous and the switches might differ significantly in terms of performance and capacity. Furthermore, the flexibility of on-demand group size makes it possible for the controller to customize its workload (\emph{e.g.}, during peak hours). As an alternative, we also implement a game-based (modified Rubinstein Bargain Model) dynamic group size limit negotiation approach in LazyCtrl. Before the controller calculates the grouping, the switches are allowed to dynamically “bargain” the group size limit with the controller according to their real-time monitored and self-evaluated data.

\end{document}